\documentclass[aps,prc,twocolumn,showpacs,superscriptaddress,floatfix,raggedbottom]{revtex4-1}
\usepackage{amsmath, amsthm, amssymb}
\usepackage{graphics}
\usepackage{graphicx}
\usepackage{epsfig}
\usepackage{gensymb}        
\usepackage{float}
\usepackage{multirow}   
\usepackage{subfig}    
\usepackage{array}

\begin{document}
\title{Analysis of the one-neutron transfer to $^{16}$O, $^{28}$Si and $^{64}$Ni induced by ($^{18}$O, $^{17}$O) reaction at 84 MeV}
\author{R. Linares}
\affiliation{Instituto de F\'isica, Universidade Federal Fluminense, 24210-340, Niter\'oi, Rio de Janeiro, Brazil}
\author{M. J. Ermamatov}
\affiliation{Instituto de F\'isica, Universidade Federal Fluminense, 24210-340, Niter\'oi, Rio de Janeiro, Brazil}
\affiliation{Institute of Nuclear Physics, Ulughbek, Tashkent, 100214, Uzbekistan}
\author{J. Lubian}
\affiliation{Instituto de F\'isica, Universidade Federal Fluminense, 24210-340, Niter\'oi, Rio de Janeiro, Brazil}
\author{F. Cappuzzello}
\affiliation{Istituto Nazionale di Fisica Nucleare, Laboratori Nazionali del Sud, I-95125 Catania, Italy}
\affiliation{Dipartimento di Fisica e Astronomia, Universit\`a di Catania, I-95125 Catania, Italy}
\author{D. Carbone}
\affiliation{Istituto Nazionale di Fisica Nucleare, Laboratori Nazionali del Sud, I-95125 Catania, Italy}
\author{E. N. Cardozo}
\affiliation{Instituto de F\'isica, Universidade Federal Fluminense, 24210-340, Niter\'oi, Rio de Janeiro, Brazil}
\author{M. Cavallaro}
\affiliation{Istituto Nazionale di Fisica Nucleare, Laboratori Nazionali del Sud, I-95125 Catania, Italy}
\author{J. L. Ferreira}
\affiliation{Instituto de F\'isica, Universidade Federal Fluminense, 24210-340, Niter\'oi, Rio de Janeiro, Brazil}
\author{A. Foti}
\affiliation{Dipartimento di Fisica e Astronomia, Universit\`a di Catania, I-95125 Catania, Italy}
\author{A. Gargano}
\affiliation{Istituto Nazionale di Fisica Nucleare, Sezione di Napoli, I-80126 Napoli, Italy}
\author{B. Paes}
\affiliation{Instituto de F\'isica, Universidade Federal Fluminense, 24210-340, Niter\'oi, Rio de Janeiro, Brazil}
\author{G. Santagati}
\affiliation{Istituto Nazionale di Fisica Nucleare, Laboratori Nazionali del Sud, I-95125 Catania, Italy}
\author{V. A. B. Zagatto}
\affiliation{Instituto de F\'isica, Universidade Federal Fluminense, 24210-340, Niter\'oi, Rio de Janeiro, Brazil}

\date{\today}

\begin{abstract} 

\textbf{Background:} Recently, a systematic exploration of two-neutron transfer induced by the  ($^{18}$O, $^{16}$O) reaction on different targets has been performed. The high resolution data have been collected at the MAGNEX magnetic spectrometer of the INFN-LNS laboratory in Catania and analyzed with the coupled reaction channel (CRC) approach. The simultaneous and sequential transfers of the two neutrons have been considered under the same theoretical framework without the need of adjustable factors in the calculations. 

\textbf{Purpose:} A detailed analysis of the one-neutron transfer cross sections is important to study the sequential two-neutron transfer. Here, we examine the ($^{18}$O, $^{17}$O) reaction on $^{16}$O, $^{28}$Si and $^{64}$Ni targets. These even-even nuclei allow for investigation of one-neutron transfer in distinct nuclear shell spaces.

\textbf{Method:} The MAGNEX spectrometer was used to measure mass spectra of ejectiles and extract differential cross sections of one-neutron transfer to low-lying states. We adopted the same CRC formalism used in the sequential two-neutron transfer, including relevant channels and using spectroscopic amplitudes obtained from shell model calculations. We also compare with one-step distorted wave Born approximation (DWBA). 

\textbf{Results:} For the $^{18}$O + $^{16}$O and the $^{18}$O + $^{28}$O systems we used two interactions in the shell model. The experimental angular distributions are reasonably well reproduced by the CRC calculations. In the $^{18}$O + $^{64}$Ni system, we considered only one interaction and the theoretical curve describes the shape and order of magnitude observed in the  experimental data.

\textbf{Conclusions:} Comparisons between experimental, DWBA and CRC angle-integrated cross sections suggest that excitations before or after the transfer of neutron is relevant in the $^{18}$O + $^{16}$O and $^{18}$O + $^{64}$Ni systems.  

\end{abstract}

\maketitle

\section{\label{Intro}Introduction}

Transfer reactions are very sensitive tools to study correlation between the initial and final nuclear states of two neighboring nuclei, providing hints to particle configurations of such states. In recent works we have dedicated our experimental and theoretical efforts to study pairing correlations in heavy-ion induced transfer reactions. Using the ($^{18}$O,$^{16}$O) reaction we have measured the two-neutron transfer cross sections leading to states in $^{14}$C \cite{CCB-13}, $^{15}$C \cite{CFC-17,CCC-15,CAA-16}, $^{18}$O  \cite{ELL-17,ECL-16}, $^{30}$Si \cite{CLL-18} and $^{66}$Ni \cite{PSV-17}. The combination of direct reaction and nuclear structure calculations have allowed to assess the role of sequential and simultaneous processes in these two-neutron transfer reactions. The sequential transfer corresponds to uncorrelated transfers of the two nucleons whereas, in the simultaneous, the particles are transferred as a single entity to the target nucleus. Comparisons between theoretical and experimental cross sections indicate that the simultaneous process is dominant in ground-to-ground two-neutron transfer reactions.

A wider picture of the sequential process in two-neutron transfer can be pursuit by measuring the one-neutron transfer reaction to the intermediate nuclear system. A successful description of this reaction channel, using the same theoretical approach, is an important evaluation for the robustness of the sequential two-neutron transfer calculations. 

In the past fifty years, many experimental studies have been performed using the (d,p) or (p,d) reactions \cite{Hod-71,Sat-83}. Recently these probes have been used to investigate single particle configurations far from stability thanks to the use of radioactive beams in inverse kinematics detector setup \cite{JAB-10,CNC-17}. Conclusions from these data were mainly established based on the distorted wave Born approximation (DWBA) grounds using effective nucleon-nucleus (NN) interactions. High-degree of accuracy in the theoretical transfer cross sections is a challenging task and requires the inclusion of deuteron breakup and the nonlocality of the NN interaction. The former can be treated within the Continuum Discretized Coupled Channel (CDCC) \cite{KYI-86}, the Alt-Grassberger-Sandhas method \cite{AGS-67} or the Adiabatic Distorted Wave Approximation (ADWA) \cite{JoT-74}. The nonlocality can be neglected in favor of a local equivalent potential with energy dependency, although deviations of about 40$\%$ compared with local potentials have been reported in Ref.~\cite{RTN-15}. Within the ADWA framework, Tsang et al. reanalyze most of the (d,p) neutron-transfer reactions using global potentials to describe the elastic scattering data of deuteron and proton \cite{TLL-05}. The obtained spectroscopic factors are remarkably consistent with large-basis shell-model calculations for sd-shell nuclei. 

In heavy-ion induced transfer, the theoretical scenario is also challenging. Firstly, effects of strong absorption are more pronounced and the angular distributions exhibit a diffraction-like pattern as the bombarding energy increases. Secondly, second-order mechanisms, such as projectile/target excitation preceding and/or following the transfer of nucleons, must be taken into account properly \cite{MGK-79}. In addition, partial waves that mostly contribute to transfer reactions are limited within a range of optimum Q-values and angular momentum transfer $l_{\textnormal{opt}}$, for a given reaction and energy \cite{Bri-72}. The one-neutron transfer reactions induced by heavy-ions have been studied recently in \cite{AGC-18,CCC-11,CSP-11}. In particular, in \cite{AGC-18}, integrated cross sections for one- and two-neutron transfers to $^{11}$B, $^{12}$C, $^{13}$C and $^{28}$Si show that the probability for two-neutron transfer is not a simple composition of two independent one-neutron transfers due to the pairing correlation between the two neutrons. This interpretation is also corroborated by results using the constrained molecular dynamics approach to describe the collision \cite{PMB-01}. 

In this work, we present a systematic analysis of the one-neutron transfer reaction to $^{16}$O, $^{28}$Si and $^{64}$Ni nuclei induced by the ($^{18}$O,$^{17}$O) reaction in the context of the coupled reaction channel (CRC) theory using a systematic optical potential. Relevant spectroscopic amplitudes in the CRC are derived from shell-model calculations. This approach provides a detailed description of the one-neutron transfer compared to the constrained molecular dynamics one and is the same adopted in the calculations of the sequential two-neutron transfer in previous works \cite{CCB-13,CFC-17,CCC-15,CAA-16,ELL-17,ECL-16,CLL-18,PSV-17}. All experimental data have been collected at the same bombarding energy (E$_\textnormal{lab}$ = 84 MeV). 

The paper is organized as follows. In Sect.~\ref{exp} the experimental details are given. Sect. III is devoted to the analysis of the experimental data using different theoretical approaches for the cross-section calculations. Results and discussions for each system are presented in Sect. IV. Finally, in Sect. V, summary and conclusions are given. 


\section{\label{exp}Experimental details}

\begin{figure}[tbp]
\centering
\graphicspath{{}}
\includegraphics[width=0.45\textwidth]{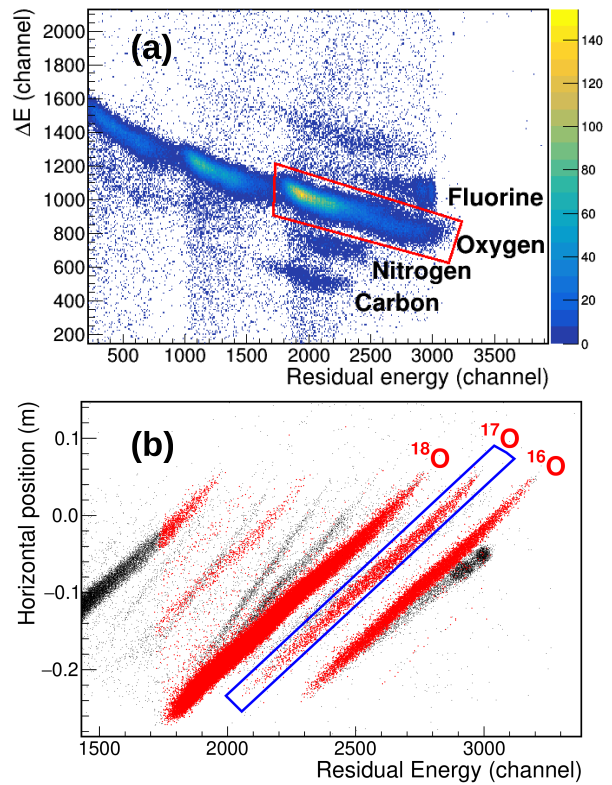}
\caption{(Color online) Typical spectra for particle identification performed at the Focal Plane Detector of the MAGNEX spectrometer. Atomic number of ejectiles are selected in a $\Delta E-E$ spectra (see Fig.~\ref{PID}a). The isotopic identification is performed exploring the correlation between horizontal position and residual energy (see Fig.~\ref{PID}b). A graphical selection on oxygen, in Fig.~\ref{PID}a, removes other atomic species and the $^{18,17,16}$O$^{8+}$ (identified in Fig.~\ref{PID}b) and a small fraction of $^{18,17,16}$O$^{7+}$ isotopes are clearly seen in red.} 
\label{PID}
\end{figure}

The measurements were performed at the {\it Istituto Nazionale di Fisica Nucleare - Laboratori Nazionali del Sud}, Catania, Italy. The 84 MeV $^{18}$O$^{6+}$ beam was delivered by the Tandem accelerator. A WO$_{3}$ ($210\pm20~\mu\textnormal{g/cm}^{2}$ thickness), $^{28}$Si ($140\pm10$ $\mu\textnormal{g/cm}^{2}$ thickness) and an enriched self-supporting $^{64}$Ni foil ($110\pm10$ $\mu\textnormal{g/cm}^{2}$ thickness) were used as targets. The $^{17}$O$^{8+}$ reaction ejectiles were momentum analyzed by the MAGNEX magnetic spectrometer \cite{CAC-16,LCC-08,LCC-09} set in full acceptance mode ($\Omega\approx50\textnormal{msr}$). The parameters of the ions trajectory (i.e. vertical and horizontal positions and incident angles) are measured by the focal plane detector, which also allows for particle identification \cite{CCC-11b}. Examples of the particle identification are shown in Figs.~\ref{PID}a and ~\ref{PID}b, exploring the $\Delta E-E$ correlation for \emph{Z} identification (Fig.~\ref{PID}a) and the horizontal position to residual energy correlation for mass selection (Fig.~\ref{PID}b). After a graphical selection of the oxygen species in the $\Delta E-E$ plot (Fig.~\ref{PID}a) only  the $^{18,17,16}$O$^{8+}$ and a small fraction of $^{18,17,16}$O$^{7+}$ isotopes are clearly seen in red (Fig.~\ref{PID}b). Note that this procedure removes interference from other nuclei like the $^{20}$F$^{9+}$, at the right of the $^{16}$O$^{8+}$ (in Fig.~\ref{PID}b). The one-neutron transfer reaction channel corresponds to a graphical selection in $^{17}$O$^{8+}$.

Trajectory reconstruction of $^{17}$O particles is performed solving the equation of motion for each detected particle \cite{LCC-08} to obtain scattering parameters at the target reference system. Further details of the data reduction technique can be found in Refs.~\cite{CAB-14,Car-15}. For the $^{16}$O($^{18}$O,$^{17}$O)$^{17}$O and the $^{28}$Si($^{18}$O,$^{17}$O)$^{29}$Si reactions, two angular settings were explored with the spectrometer optical axis centered at 8$\degree$ and 10$\degree$. Due to the large angular acceptance of the spectrometer, these angular settings correspond to a total covered angular range of $4\degree < \theta_{\textnormal{lab}} < 15\degree$, with an overlap of $~8\degree$ between the two angular settings. For the $^{64}$Ni($^{18}$O,$^{17}$O)$^{65}$Ni reaction, three angular settings were explored, with the spectrometer optical axis centered at $\theta_{\textnormal{lab}}$ = 12$\degree$, 18$\degree$ and 24$\degree$. For this reaction, the total covered angular range is about $7\degree < \theta_{\textnormal{lab}} < 29\degree$.


\begin{figure}[tbp]
\centering
\graphicspath{{}}
\includegraphics[width=0.45\textwidth]{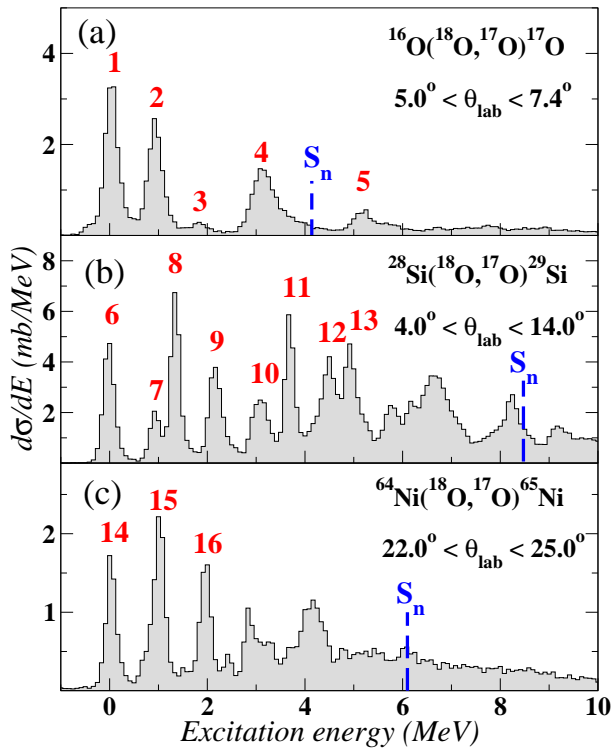}
\caption{(Color online) Excitation energy spectra of (a) $^{17}$O, (b) $^{29}$Si and (c) $^{65}$Ni residual nuclei. Backgrounds due to contamination in the targets are negligible under the peaks of interest and are not shown in the spectra. The blue lines are set at one-neutron emission threshold energy for each nucleus. Transfer cross sections have been extracted  for the numbered peaks indicated in each spectrum.} 
\label{spectrum}
\end{figure}

The excitation energy spectra, relative to the ground to ground states (g.s.) transition Q-value for each reaction, are shown in Fig.~\ref{spectrum}. The overall energy resolution is about 250 keV (Full Width at Half Maximum). A list of the main peaks identified in Fig.~\ref{spectrum} is presented in Table~\ref{table:ListOfStates}. A supplementary measurement was performed using a $50\pm5$ $\mu\textnormal{g/cm}^{2}$ self-supporting $^{12}$C target for the estimation of the contribution from carbon backing on the WO$_3$ target. In this case, contamination was negligible and it is not shown in the spectrum (Fig.~\ref{spectrum}a). Contamination from carbon build-up in the $^{28}$Si and $^{64}$Ni targets does not interfere with the analyzed peaks.

\begin{table*} [t]
\caption{List of the main states of $^{17}$O, $^{29}$Si and $^{65}$Ni nuclei populated with the one-neutron transfer. Label of the peaks are indicated in Fig. 1.} 
\centering
\begin{tabular}{c c c c | c c c c | c c c c} 
\hline
\multicolumn{4}{c}{\textbf{$^{17}$O}}  & \multicolumn{4}{c}{\textbf{$^{29}$Si}} & \multicolumn{4}{c}{\textbf{$^{65}$Ni}} \\ \hline

\textbf{label} & \textbf{E$_\textbf{exp}$} & \textbf{E$_\textbf{ref}$} & \textbf{J$^{\pi}$} & \textbf{label} & \textbf{E$_\textbf{exp}$} & \textbf{E$_\textbf{ref}$} & \textbf{J$^{\pi}$} & \textbf{label} & \textbf{E$_\textbf{exp}$} & \textbf{E$_\textbf{ref}$} & \textbf{J$^{\pi}$} \\
         &     (MeV)           &   (MeV)    &   &         &     (MeV)           &   (MeV)    &   &         &     (MeV)           &   (MeV)    &   \\ 
\hline
\hline
                       
 1 & 0.00 & g.s.  & $5/2^{+}$ & 6 & 0.00 & g.s.  & $1/2^{+}$ & 14 & 0.05 & g.s  & $5/2^{-}$  \\ 
   &      &      &           &   &      &      &           &    &      & 0.06 & $1/2^{-}$  \\
   &      &      &           &   &      &      &           &    &      & 0.31 & $3/2^{-}$  \\ \hline
   
 2 & 0.87$^{b}$ & 0.87 &  $1/2^{+}$& 7 & 0.88$^{a}$& &           & 15 & 0.98$^{b}$ & 0.69 & $3/2^{-}$  \\
   &      &      &           &   &      &      &           &    &      & 1.02 & $9/2^{+}$  \\
   &      &      &           &   &      &      &           &    &      & 1.14 & ($7/2^{-}$)  \\
   &      &      &           &   &      &      &           &    &      & 1.27 & $5/2^{-}$  \\ \hline  
   
 3 & 1.74$^{c}$ & - & -      & 8 & 1.30 & 1.27 & $3/2^{+}$ & 16 & 1.96$^{b}$& 1.92 &$5/2^{+}$  \\ \hline
 
 4 & 3.15 & 3.06 & $1/2^{-}$ & 9 & 2.09 & 2.03 & $5/2^{+}$ &    &      &      &            \\ \hline
 
 5 & 5.20 & 5.08 & $3/2^{+}$ & 10& 3.05 & 3.07 & $5/2^{+}$ &    &      &      &            \\ 
   &      & 5.13 & $9/2^{-}$ &   &      &      &           &    &      &      &            \\ 
   &      & 5.38 & $3/2^{-}$ &   &      &      &           &    &      &      &            \\ \hline
   
   &      &      &           & 11& 3.64 & 3.62 & $7/2^{-}$ &    &      &      &            \\ \hline
   
   &      &      &           & 12& 4.46$^{b}$& - & -       &    &      &      &            \\ \hline
   
   &      &      &           & 13& 4.89 & 4.74 &($9/2^{+}$)&    &      &      &            \\
   &      &      &           &   &      & 4.84 & $1/2^{+}$ &    &      &      &            \\
   &      &      &           &   &      & 4.90 & $5/2^{+}$ &    &      &      &            \\  
   &      &      &           &   &      & 4.93 & $3/2^{-}$ &    &      &      &            \\ \hline
\end{tabular}

\raggedright
$^{a}$Peak corresponds to projectile excitation. 

$^{b}$Peaks that may also contain projectile excitation along with states of the residual nuclei.

$^{c}$Peak corresponds to sum of the projectile and target excitations. 
\label{table:ListOfStates}
\end{table*} 

For the $^{16}$O($^{18}$O,$^{17}$O)$^{17}$O reaction, the experimental energy resolution allows for a clear identification of the ground and $1/2^{+}$ ($0.87~\textnormal{MeV}$) states. The peaks corresponding to these two states are numbered as 1 and 2 in Fig.~\ref{spectrum}a, respectively. Since the residual and ejectile particles are the same in the outgoing mass partition, the peak 2 may also contains a non-negligible contribution from excitation of the ejectile, leaving the residual nucleus in the ground state. The third peak (numbered as 3 in Fig.~\ref{spectrum}a), at $\approx 1.74$ MeV, is generated by the simultaneous excitation of the $1/2^{+}$ state in both projectile and residual nuclei, i.e. $^{16}$O($^{18}$O,$^{17}$O$_{0.87}$)$^{17}$O$_{0.87}$. 

In the $^{28}$Si($^{18}$O,$^{17}$O)$^{29}$Si reaction, the ground ($5/2^{+}$) and first excited ($3/2^{+}$) states of $^{29}$Si are identified by numbers 6 and 8, respectively (see Fig.~\ref{spectrum}b). The peak 7 corresponds to the excitation of the $^{17}$O projectile ($1/2^{+}$) at $0.87~\textnormal{MeV}$. Other peaks appear at higher excitation energies in the spectrum and correspond to the unresolved states of $^{29}$Si and/or $^{17}$O.

In the $^{64}$Ni($^{18}$O,$^{17}$O)$^{65}$Ni reaction, the first peak (label 14 in Fig.~\ref{spectrum}c) corresponds to three unresolved states of $^{65}$Ni, namely the $5/2^{-}$ (g.s.), $1/2^{-}$ ($0.06~\textnormal{MeV}$) and $3/2^{-}$ ($0.31~\textnormal{MeV}$). The second peak (label 15 in Fig.~\ref{spectrum}c) corresponds to four unresolved states, namely the $3/2^{-}$ ($0.69~\textnormal{MeV}$), $9/2^{+}$ ($1.02~\textnormal{MeV}$), $7/2^{-}$ ($1.14~\textnormal{MeV}$) and $5/2^{-}$ ($1.27~\textnormal{MeV}$).

Angular distributions for the one-neutron transfer cross sections leading to the states indicated in Fig.~\ref{spectrum} have been obtained and are shown in the next sections. The angular resolution is $0.3\degree$ for $^{64}$Ni and $0.5\degree$ for $^{16}$O and $^{28}$Si. The error bars in the cross sections correspond to an uncertainty in the solid angle determination and the statistical error. A systematic uncertainty of 10$\%$, coming from the target thickness and beam integration by the Faraday cup, is common to all the angular distribution points and it is not included in the error bars.

\section{\label{theor}Theoretical Framework}

The CRC approach is adopted for the calculation of the direct reaction cross sections. In this framework, calculations were performed using the FRESCO code \cite{Tho-88,ThN-09} with the nuclear S\~ao Paulo potential (SPP)  \cite{CCG-02} as optical potential in both mass partitions. States considered in each system are indicated in the next subsections. In the entrance partition, a normalization coefficient of 0.6 for the imaginary part of the SPP was used to account for dissipative processes and missing couplings to continuum states \cite{PLO-09,PLO-12}. Due to the fact that the mass diffuseness of the $^{18}$O projectile does not follow the systematics of the SPP, we used the value of 0.61 fm as indicated in Ref.~\cite{COS-11}. In the outgoing partition, the imaginary part is scaled by a larger factor (0.78). This optical potential describes the angular distribution of elastic and quasi-elastic cross sections for many systems in a wide energy interval \cite{GCG-06} and also an unexpected rainbow-like pattern in the elastic scattering of $^{16}$O + $^{27}$Al system at 100 MeV \cite{OCC-13} without the need of adjustable parameters. Moreover, this is also the same potential adopted in previous works on two-neutron transfer reaction to $^{12,13}$C, $^{16}$O, $^{28}$Si and $^{64}$Ni at the same bombarding energy.

The potentials for the calculations of one-neutron wave functions were assumed to have a Woods-Saxon shape with reduced radii and diffuseness values set at $1.2~\textnormal{fm}$ and $0.6~\textnormal{fm}$ for the nuclear cores, respectively. The depth of these shapes were varied in order to fit the experimental separation energies for one neutron. In recent studies on two-neutron transfer induced by ($^{18}$O,$^{16}$O) reaction \cite{CCB-13,CFC-17,CCC-15,CAA-16,ELL-17,ECL-16,CLL-18,PSV-17}, this approach has proven to be adequate for lighter nuclei, such as $^{12}$C and $^{16}$O nuclei. In the $^{64}$Ni($^{18}$O,$^{16}$O)$^{66}$Ni reaction, the sequential two-neutron transfer provides a better agreement with experimental data for the $\textnormal{g.s.} \rightarrow 2^{+}$ transition using slightly different values for the reduced radii and diffuseness parameters that were obtained with the interacting boson-fermion model \cite{IaS-91} to describe the wave functions for states in $^{64,65}$Ni \cite{PSV-17}. Nevertheless, for one-neutron transfer, we found that this dependence is not as critical as in the sequential two-neutron transfer and we set the same values for the $^{64}$Ni core. 

\begin{figure}[tbp]
\centering
\graphicspath{{}}
\includegraphics[width=0.45\textwidth]{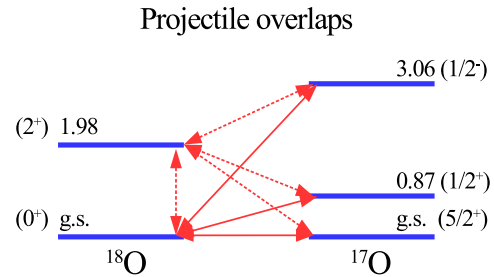}
\caption{(Color online) Coupling scheme for projectile overlaps considered in this work. Continous and dashed lines correspond to direct and two-step transitions. } 
\label{Coupling}
\end{figure}

Spectroscopic amplitudes (SA) were extracted from the shell model calculations using the NuShellX code \cite{Nushell}. The coupling scheme for the projectile overlaps, used throughout this work, is shown in Fig.~\ref{Coupling}. Similar schemes were considered in the previous analyses of sequential two-neutron transfers to $^{12}$C \cite{CCB-13}, $^{13}$C \cite{CFC-17}, $^{16}$O \cite{ECL-16} and $^{64}$Ni \cite{PSV-17}. The inclusion of high-lying excited states in $^{18}$O or $^{17}$O nuclei does not change appreciably the  cross sections for the one-neutron transfers. The target overlaps for each nucleus are described in the following section.

\section{\label{disc}Results and Discussions}

The main purpose here is to treat three different one-neutron transfer reactions under the same theoretical framework, avoiding arbitrary normalization factors in the calculations. In this sense, the $^{16}$O($^{18}$O,$^{17}$O)$^{17}$O reaction provides a good workbench to assess nuclear structure models of neighboring stable isotopes of oxygen. Such information is required to describe the one-neutron transfer cross sections in the other two systems. Different model spaces and interactions, suitable for $^{28,29}$Si and $^{64,65}$Ni target nuclei, are used in the calculations of SA required for the $^{28}$Si($^{18}$O,$^{17}$O)$^{29}$Si and $^{64}$Ni($^{18}$O,$^{17}$O)$^{65}$Ni reactions, respectively.

\subsection{\label{o-16}Transfer to the $^{16}$O nucleus}

Two interactions are typically used in shell model calculations in this mass region: the Zucker-Buck-McGrory (ZBM)  \cite{ZBM-68} and the \textit{psdmod}, that is a modified version of the \textit{psdwbt} interaction \cite{UtC-11}. In the first model, a $^{12}$C core is considered and the valence subspace includes the 1p$_{1/2}$, 1d$_{5/2}$, 2s$_{1/2}$ orbitals. In the second one, a $^{4}$He core is adopted and the valence subspace is 1p$_{1/2}$, 1p$_{3/2}$, 1d$_{3/2}$, 1d$_{5/2}$, 2s$_{1/2}$, described by the potential of Ref.~\cite{UtC-11}. These two interactions were considered to study the two-neutron transfer in the $^{16}$O($^{18}$O,$^{16}$O)$^{18}$O, $^{16}$O(t,p)$^{18}$O \cite{ELL-17,ECL-16} and $^{13}$C($^{18}$O,$^{16}$O)$^{14}$C reactions \cite{CFC-17}. From these works, the ZBM interaction describes the experimental absolute cross sections of the two-neutron transfer reasonably well. Despite of that, we still consider the \textit{psdmod} interaction to compare both interactions in the context of one-neutron transfer. In the target nucleus, the $0^{+}$ (g.s.) and the $3^{-}$ (6.05 MeV) of the $^{16}$O nucleus were considered in the coupling scheme of the CRC calculations along with the overlaps with the first three excited states of the $^{17}$O nucleus. The spectroscopic amplitudes for these overlaps can be found in Ref.~\cite{CAA-16}.

\begin{figure}[tbp]
\centering
\graphicspath{{}}
\includegraphics[width=0.5\textwidth]{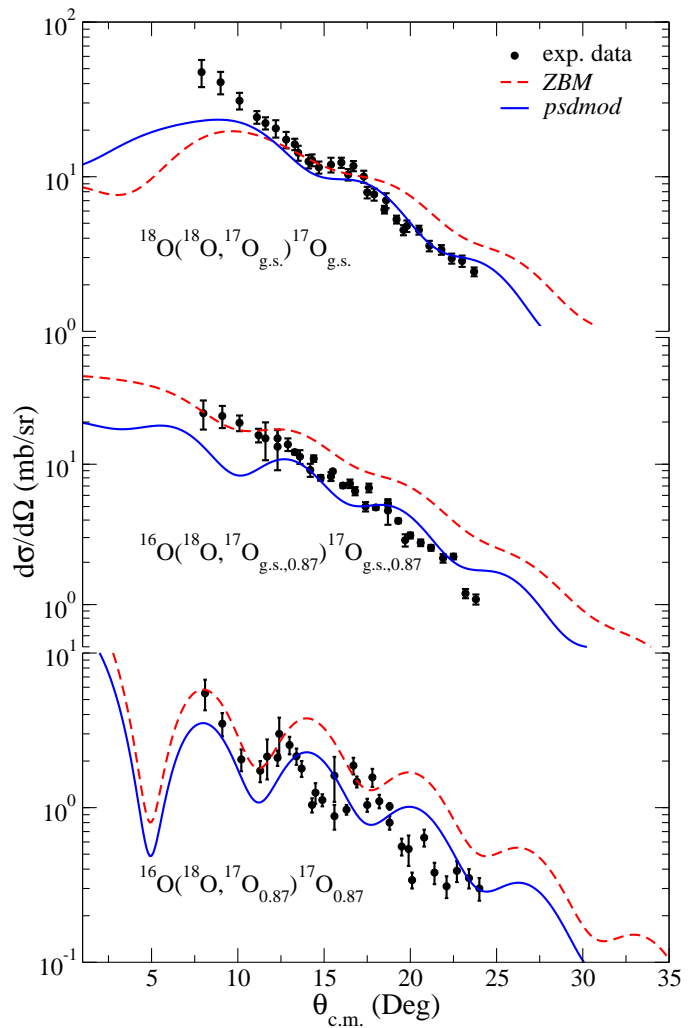}
\caption{(Color online) Experimental angular distribution of the cross sections for the one-neutron transfer in the $^{16}$O($^{18}$O,$^{17}$O)$^{17}$O reaction compared with theoretical predictions. The reaction channels are identified in the graph. Solid blue and dashed red curves correspond to CRC calculations using SA derived with the ZBM and \textit{psdmod} interactions, respectively.
} 
\label{results_o16a}
\end{figure}

The experimental angular distribution for one-neutron transfer leading to the ground and $1/2^{+}$ state of $^{17}$O nucleus is shown in Fig.~\ref{results_o16a}. As it was already mentioned in Sect.~\ref{Intro}, the reaction results identical particles in the outgoing mass partition, both $^{17}$O particles in the ground state and in the first excited state. For these cases, theoretical curves correspond to the coherent sum of scattering amplitudes at $\theta$ and $(\pi - \theta)$. Another point to be stressed is that the experimental setup does not distinguish between the $^{16}$O($^{18}$O,$^{17}$O$_{\textnormal{g.s.}}$)$^{17}$O$_{0.87}$ and $^{16}$O($^{18}$O,$^{17}$O$_{0.87}$)$^{17}$O$_{\textnormal{g.s.}}$ reaction channels because they have the same Q-value. In this case, experimental cross sections were determined from counting events under peak 2 (see Fig.~\ref{spectrum}a) and data are shown in Fig.~\ref{results_o16a} along with CRC calculations summed for the two channels that compose this peak. 

Results using the \textit{psdmod} interaction to derive the SAs show an overall good agreement with experimental data. The ZBM overestimates the cross sections even though also describe the shape of experimental data. In the analysis of other two system we are considering the \textit{psdmod} for the projectile overlaps.

\subsection{\label{si-28}Transfer to the $^{28}$Si nucleus}

In \cite{CLL-18} we have observed that deformation of the target nucleus results in a competition between sequential and simultaneous two-neutron transfer in $^{28}$Si. The nuclear surface of the ground state of $^{28}$Si is not spherical (reduced electric quadrupole transition probability is $0.0326\pm0.0012$ $\textnormal{e}^{2}\textnormal{b}^{2}$ \cite{RNT-01}). Mermaz and co-workers have measured the angular distribution of several multinucleon transfer reactions to $^{28}$Si induced by $^{18}$O at 56 MeV \cite{MGK-79}. Single- and multi-step mechanisms were considered under the scope of the DWBA and the coupled channel Born approximation (CCBA) approaches, respectively. The authors needed to arbitrary decrease the effective Coulomb barrier at the surface in order to reproduce the elastic and inelastic channels, as well as the shape of angular distributions of the transfer cross sections,  Peterson et al.~\cite{PFR-83} studied the (d,p) reaction on $^{28}$Si and adopted the weak-coupling model to treat the $^{29}$Si as a valence neutron in the 2s-1d shells coupled to the collective states of $^{28}$Si core. Calculations indicated that the one-neutron transfer to the $7/2^{+}$ (4.08 MeV) in the $^{29}$Si proceeds mainly through the lowest collective $2^{+}$ state of $^{28}$Si. Moreover, it was shown that this second-order mechanism is negligible for the population of the $1/2^{+}$ (g.s.) and $3/2^{+}$ (1.27 MeV) states of $^{29}$Si.

In the present work, spectroscopic information for the states in the $^{28,29}$Si nuclei are extracted from two different phenomenological interactions: the effective \textit{psdmod} and the \textit{psdmwkpn} interaction \cite{BaD-13}. The latter is a combination of the Cohen-Kurath interaction \cite{CoK-65} for the p-shell, the Wildenthal interaction \cite{Wil-84} for the sd-shell and the Millener-Kurath interaction \cite{MiK-75} for the coupling matrix elements between p- and sd-shells. 

Within both interactions, the model space assumes $^{4}$He as a closed core and valence neutrons and protons in the 1p$_{3/2}$, 1p$_{1/2}$, 1d$_{3/2}$, 1d$_{5/2}$, and 2s$_{1/2}$ orbitals. This model space is larger compared to the one considered by Peterson and co-workers~\cite{PFR-83}. Coupling between single-particle states in the 2s-1d shells and the $^{28}$Si core, which was important to describe the (d,p) data in Ref.~\cite{PFR-83}, is effectively considered within our model space.

The SAs derived from the \textit{psdmod} and \textit{psdmwkpn} interactions are listed in Table~\ref{amplitudes_si-29}. The SAs values are very close each other with some deviations, like in the SA for $^{28}$Si$_ {g.s.}$ to $^{29}$Si$_ {g.s.}$ transition. In the last four columns of Table~\ref{amplitudes_si-29} we also compare spectroscopic factors (SF), obtained in this work, with those found by DWBA analysis in Refs.~\cite{MGK-79,PFR-83}.  The SF for the g.s. to g.s. transition from \textit{psdmod} interaction is close to the value obtained by Mermaz et al.~\cite{MGK-79} whereas the \textit{psdmwkpn} agrees with the value reported by Peterson et al.~\cite{PFR-83}. 

\begin{table*} [t]
\caption{One-neutron spectroscopic amplitudes (SA) and spectroscopic factors (SF) for $^{28}$Si to $^{29}$Si transitions obtained by shell model calculations using \textit{psdmod} and \textit{psdmwkpn} interactions. $nlj$ are the principal quantum numbers, the orbital and the total angular momenta of the single neutron. We also include SFs deduced by (d,p) reactions found in Refs.~\cite{MGK-79,PFR-83}.} 
\centering
\begin{tabular}{ c|c|c|c c|c c c c }
\hline
   &   &   & \multicolumn{2}{c}{\textbf{SA}} & \multicolumn{4}{c}{\textbf{SF}} \\
\textbf{Initial State} & \textbf{nl$_j$} & \textbf{Final State} & \textit{psdmod} & \textit{psdmwkpn} & \textit{psdmod} & \textit{psdmwkpn} & \text{Ref.~\cite{MGK-79}} & \text{Ref.~\cite{PFR-83}} \\ \hline
\hline

$^{28}$Si$_{\textnormal{g.s.}}$ & $(2s_{1/2})$  & $^{29}$Si$_{g.s.} (1/2^+)$   & +0.716 & +0.570 & 0.51 & 0.32 & 0.53 & 0.37 \\ 
($0^{+}$) & $(1d_{3/2})$  & $^{29}$Si$_{1.27} (3/2^+)$   & -0.827 & -0.806 & 0.68 & 0.65 & 0.74 & 0.53 \\ 
        & $(1d_{5/2})$  & $^{29}$Si$_{2.03} (5/2^+)$   & -0.347 & -0.451 & 0.12 & 0.20 & 0.12 & 0.08 \\
        & $(1d_{3/2})$  & $^{29}$Si$_{2.43} (3/2^+)$   & +0.046 & +0.007 & 0.002&  -   &   -  &  -   \\
        & $(1d_{5/2})$  & $^{29}$Si$_{3.07} (5/2^+)$   & -0.226 & -0.247 & 0.05 & 0.06 & 0.06 & 0.03 \\ \hline
        
$^{28}$Si$_{1.78}$ & $(1d_{3/2})$  & $^{29}$Si$_{g.s.} (1/2^+)$   & -0.388 & -0.479 &   &   &   &   \\
($2^{+}$) & $(1d_{5/2})$  &   & -0.847 & -0.857 &   &   &   &   \\
        & $(2s_{1/2})$  & $^{29}$Si$_{1.27} (3/2^+)$  & -0.090 & -0.019 &   &   &   &   \\
        & $(1d_{3/2})$  &                             & -0.006 & +0.027 &   &   &   &   \\
        & $(1d_{5/2})$  &                             & +0.293 & +0.345 &   &   &   &   \\
        & $(2s_{1/2})$  & $^{29}$Si$_{2.03} (5/2^+)$  & +0.632 & +0.562 &   &   &   &   \\
        & $(1d_{3/2})$  &                             & +0.025 & -0.037 &   &   &   &   \\
        & $(1d_{5/2})$  &                             & +0.414 & +0.478 &   &   &   &   \\
        & $(2s_{1/2})$  & $^{29}$Si$_{2.43} (3/2^+)$  & -0.341 & -0.247 &   &   &   &   \\
        & $(1d_{3/2})$  &                             & +0.748 & +0.764 &   &   &   &   \\
        & $(1d_{5/2})$  &                             & -0.518 & -0.587 &   &   &   &   \\
        & $(2s_{1/2})$  & $^{29}$Si$_{3.07} (5/2^+)$  & +0.013 & -0.003 &   &   &   &   \\
        & $(1d_{3/2})$  &                             & -0.761 & -0.742 &   &   &   &   \\
        & $(1d_{5/2})$  &                             & +0.052 & +0.034 &   &   &   &   \\ \hline
        
$^{28}$Si$_{4.62}$ & $(1d_{3/2})$  & $^{29}$Si$_{1.27} (3/2^+)$   & +0.904 & +0.645 &   &   &   &   \\
($4^{+}$) & $(1d_{3/2})$  & $^{29}$Si$_{2.03} (5/2^+)$  & -0.425 & +0.452 &   &   &   &   \\
        & $(1d_{5/2})$  &                             & -0.195 & +0.201 &   &   &   &   \\
        & $(1d_{5/2})$  & $^{29}$Si$_{2.43} (3/2^+)$  & +0.295 & +0.309 &   &   &   &   \\
        & $(1d_{3/2})$  & $^{29}$Si$_{3.07} (5/2^+)$  & -0.244 & -0.074 &   &   &   &   \\
        & $(1d_{5/2})$  &                             & +0.406 & +0.666 &   &   &   &   \\ \hline

\end{tabular}
\label{amplitudes_si-29}
\end{table*}

\begin{figure}[tbp]
\centering
\graphicspath{{}}
\includegraphics[width=0.45\textwidth]{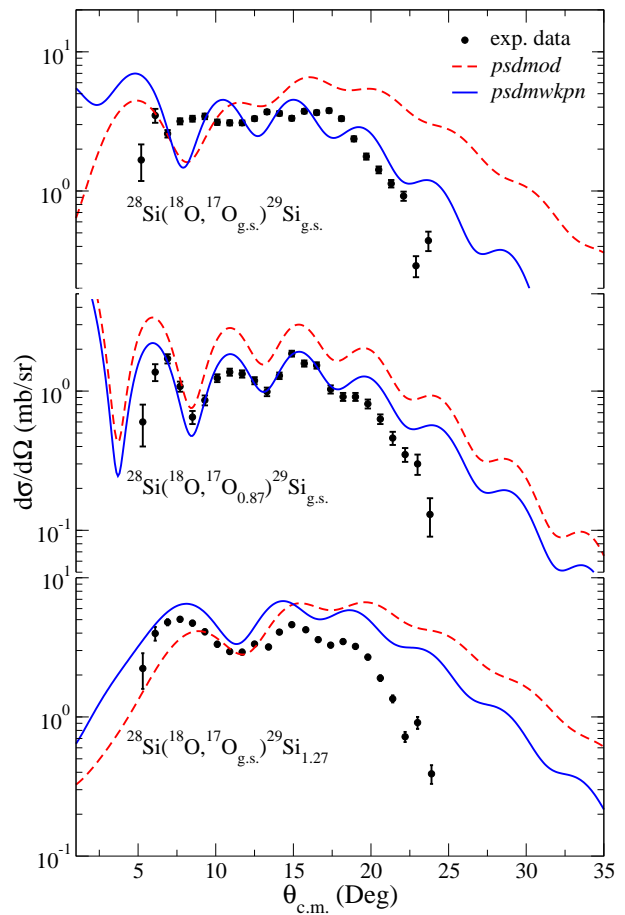}
\caption{(Color online) Angular distributions of the cross sections for the $^{28}$Si($^{18}$O,$^{17}$O)$^{29}$Si reaction leading to the population of $^{28}$Si in the ground state (top); (c) $^{17}$O ejectile in the $1/2^{+}$ state at 0.87 MeV (middle) and $^{28}$Si in the $3/2^{+}$ state at 1.27 MeV (bottom). SAs for $^{17,18}$O derived in the shell model using the \textit{psdmod}. In the $^{28,29}$Si cases, two interactions have been considered: the \textit{psdmod} and the \textit{psdmwkpn}.} 
\label{results_si28}
\end{figure}

The comparison between the experimental and theoretical angular distributions of the cross-sections for one-neutron transfer reaction using the \textit{psdmod} and \textit{psdmwkpn} interactions is shown in Fig.~\ref{results_si28}. The overall best description of the experimental data is achieved using SAs from the {\it psdmwkpn} interaction. For the $^{28}$Si($^{18}$O,$^{17}$O$_{g.s.}$)$^{29}$Si$_{g.s.}$ reaction channel (Fig.~\ref{results_si28}a) exhibit a slightly poor agreement between theoretical curves and experimental data compared to other reaction channels (Figs.~\ref{results_si28}b and ~\ref{results_si28}c) although the two theoretical curves give cross section of the same order of magnitude as the experimental one. For transitions to the $3/2^{+}$ state (1.27 MeV) in $^{29}$Si, the local minimum at $13\degree$ (c.m.) is reproduced in both calculations (see Fig.~\ref{results_si28}b), although calculations with the \textit{psdmod} interaction overestimate the cross sections at $\theta_{CM} > 15\degree$. In the $^{28}$Si($^{18}$O,$^{17}$O$_{0.87}$)$^{29}$Si$_{g.s.}$ reaction channel, the experimental data are very well described by the \textit{psdmwkpn} interaction (Fig.~\ref{results_si28}c) while the \textit{psdmod} calculations are well above the data.

A survey among other microscopic interactions for the $^{28,29}$Si nuclei has been considered but that has led to poor agreement with experimental data and are not shown here. From a simple comparative inspection, the low-quality agreement observed between theory and experimental data for the ground-to-ground transition to $^{28}$Si nucleus seems to be related to its deformation since a better agreement is observed for the spherical $^{16}$O nucleus. On the other hand, quite good agreement have been obtained to the $^{28}$Si(t,p)$^{30}$Si reaction using the \emph{psdmod} interaction \cite{CLL-18}. In fact, measurements at very forward angle would be helpful to better judge the agreement of calculations in the present work. The effect of deformation in heavy-ion induced transfer demands further study.

\subsection{\label{ni-64}Transfer to the $^{64}$Ni nucleus}

In the literature, there are some discrepancies between data and theoretical one-neutron SF for the Ni nuclei. Lee et al.~\cite{LTL-09} have performed a systematic reanalysis of the angular distributions measured with (d,p) reactions on Ni isotopes using the ADWA for the reaction model. Using large-basis shell model calculations, an overall deviation of about 25$\%$ was observed with respect to the SF reported in previous works. In some cases, deviations of about 60$\%$ can be found as for the ground state of the $^{65}$Ni nucleus (see Table II in Ref.~\cite{LTL-09}).

The two-neutron transfer $^{64}$Ni($^{18}$O,$^{16}$O)$^{66}$Ni reaction has been studied in Ref.~\cite{PSV-17}. In that work, the sequential two-neutron transfer, in which transitions to states in the $^{65}$Ni nucleus are an intermediate step, is important in order to interpret the experimental two-neutron transfer cross sections that lead to the $2^{+}$ states of the $^{66}$Ni nucleus. In the shell model calculation, the neutron SAs for $^{65}$Ni were derived using the \textit{bjuff} model space. In this model space, the $^{48}$Ca is considered as a closed core with valence particles populating the 1f$_{7/2}$ and 2p$_{3/2}$ orbitals for protons and the 2p$_{3/2}$, 1f$_{5/2}$, 2p$_{1/2}$, and 1g$_{9/2}$ orbitals for neutrons. An effective interaction derived from CD-Bonn nucleon-nucleon potential was adopted~\cite{PSV-17}, developed for Nickel isotopes with mass near to $A=64$. The SA values can be found in Ref.~\cite{PSV-17}. This approach produces energy spectra of $^{64,65}$Ni isotopes in good agreement with the experimental data, within 200 keV of deviation for both negative- and positive-parity states. 

\begin{figure}[tbp]
\centering
\graphicspath{{}}
\includegraphics[width=0.45\textwidth]{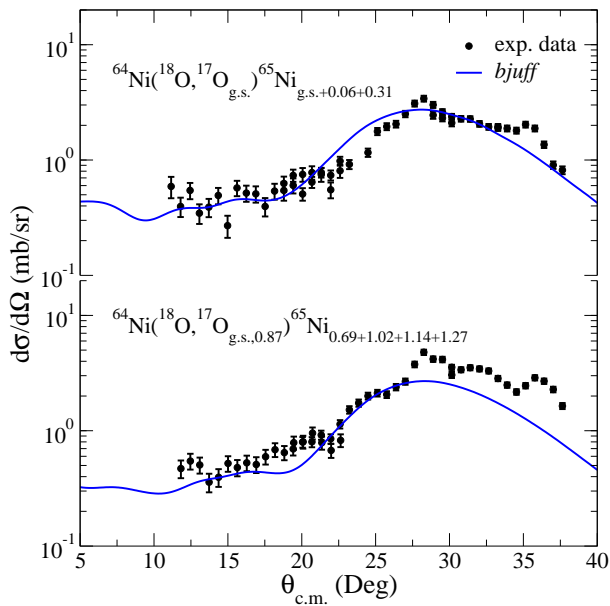}
\caption{(Color online) Angular distributions of the one-neutron transfer cross sections leading to the low-lying states in the $^{65}$Ni. At the top, summed cross sections for transfers to the g.s., 0.06 MeV and 0.31 MeV states in the $^{65}$Ni nucleus. Bottom, summed cross sections for the 0.69 MeV, 1.02 MeV, 1.14 MeV and 1.27 MeV states in $^{65}$Ni and 0.87 MeV in $^{17}$O ejectile. SAs for $^{17,18}$O and $^{64,65}$Ni derived in the shell model using \textit{psdmod} and \textit{bjuff} interactions, respectively.} 
\label{results_ni64}
\end{figure}

Experimental results for the one-neutron transfer reaction on $^{64}$Ni are shown in Fig.~\ref{results_ni64}, along with the theoretical curves using SA obtained from shell model. Experimental data points and calculations correspond to the sum of the one-neutron transfer cross sections for the ground, 0.06 MeV and 0.31 MeV states of $^{65}$Ni (Fig.~\ref{results_ni64}) and for the 0.69, 1.27 and 1.41 MeV states of the $^{65}$Ni with the $^{17}$O ejectile at 0.87 MeV (Fig.~\ref{results_ni64}). The theoretical curves reproduce quite well the bell-like shape of the angular distributions around $\theta = 30\degree$. In general, one can say that the overall agreement between the theoretical calculations and the data is reasonably good, considering that we are not fitting any parameter here.

\subsection{\label{AngleIntegrated}Angle-integrated cross sections}

Dominant factors to the angle-integrated cross sections of the one-neutron transfer are the binding energy of the transferred neutrons, the kinematic matching between the relative motion of the cores and the valence particle and the spin configuration. In the projectile, the binding energy of the valence neutron to the $^{17}$O core is a common feature to all three systems. The second factor accounts for the linear and angular momentum dependence in the entrance and exit channels. Final states that satisfy a relation between transferred angular momentum and Q-value, indicated by the Brink's rule \cite{Bri-72}, are strongly populated. Finally, the third factor is related to spin configuration of the transferred nucleon before and after the transfer reaction, in which those that take place without spin flip are more favored against reactions that requires flipping of the transferred spin \cite{FrL-96}. 

Here we compare the experimental angle-integrated cross sections ($\sigma_{exp}$) with the theoretical ones derived from DWBA ($\sigma_{DWBA}$) and CRC ($\sigma_{CRC}$) calculations for the three systems. Quantum mechanical calculations within the one-step DWBA, using the SFs derived from shell model, implicitly take the above-mentioned factors into account. Nevertheless, deviations from the data can be observed due to the second order processes, such as projectile/target excitation prior to the transfer, which may play a role for different target nuclei. These processes can be explicitly included within a CRC model, thus the comparison between experimental data with DWBA and CRC cross sections indicates the effect of multi-step processes in the nucleon transfer dynamics. 

\begin{table*} [t]
\caption{Experimental and theoretical angle-integrated cross sections (in mb) of the one-neutron transfer reactions for the final states in $^{17}$O, $^{29}$Si and $^{65}$Ni nuclei. We show angle-integrated cross sections obtained using the two interactions considered in this work within the DWBA and CRC. In the case of $^{17}$O, interaction A e B stand for \emph{psdmod} and \emph{zbm} interactions, respectively. For the $^{29}$Si case, stand for \emph{psdmwkpn} and \emph{psdmod}, respectively. Uncertainties in the experimental angle-integrated cross sections are estimated to be 10$\%$ due to systematic uncertainties in the cross sections. Angular ranges for the angle-integrated cross section are specified in the text.} 
\centering
\begin{tabular}{ c c c c c c c c c c c }
\hline
	&	&	& \multicolumn{4}{c}{\textbf{interaction A}} & \multicolumn{4}{c}{\textbf{interaction B}} \\
\textbf{system} & \textbf{channel} & $\sigma_{\textnormal{exp}}$ (mb) & $\sigma_{\textnormal{DWBA}}$ (mb) & $\sigma_{\textnormal{CRC}}$ (mb) & $\delta_{\textnormal{DWBA}}^a$ & $\delta_{\textnormal{exp}}^b$ &
$\sigma_{\textnormal{DWBA}}$ (mb) & $\sigma_{\textnormal{CRC}}$ (mb) & $\delta_{\textnormal{DWBA}}^a$ & $\delta_{\textnormal{exp}}^b$ \\
\hline
\hline

\multirow{4}{*}{$^{18}$O+$^{16}$O} & $^{17}$O$_{g.s.}$+$^{17}$O$_{g.s.}$ & 4.2 & 6.5 & 4.5 & $44\%$ & $-7\%$  & 6.8 & 5.0 & $36\%$ & $-16\%$  \\ 
                  & $\left. \begin{tabular}{@{\ }l@{}}
    $^{17}$O$_{0.87}$+$^{17}$O$_{g.s.}$ \\ $^{17}$O$_{g.s.}$+$^{17}$O$_{0.87}$ 
  \end{tabular}\right\}$  & 3.1 & 5.1 & 3.4 & $50\%$ & $-9\%$ & 7.1 & 6.0 & $18\%$ & $-48\%$ \\
				& $^{17}$O$_{0.87}$+$^{17}$O$_{0.87}$ & 0.7 & 1.0 & 0.6 & $66\%$ & $16\%$ & 1.5 & 0.9 & $67\%$ & $-22\%$  \\

\hline
\multirow{3}{*}{$^{18}$O+$^{28}$Si} & $^{17}$O$_{g.s.}$+$^{29}$Si$_{g.s.}$ & 1.3 & 1.7 & 1.5 & $13\%$ & $-13\%$  & 2.8  & 3.1 & $-10\%$ & $-58\%$ \\
				 & $^{17}$O$_{0.87}$+$^{29}$Si$_{g.s.}$ & 0.5 &  0.7   & 0.6 & $17\%$ & $-16\%$  &   1.9 & 2.2  & $-14\%$  &  $-77\%$ \\
                 & $^{17}$O$_{g.s.}$+$^{29}$Si$_{1.27}$ & 1.5 & 3.0 & 2.6 & $15\%$ & $-42\%$  &   3.3 & 3.7  & $-11\%$  &  $-59\%$ \\
\hline

\multirow{8}{*}{$^{18}$O+$^{64}$Ni} & $\left. \begin{tabular}{@{\ }l@{}} $^{17}$O$_{g.s.}$+$^{65}$Ni$_{g.s.}$ \\ 
$^{17}$O$_{g.s.}$+$^{65}$Ni$_{0.06}$ \\ 
$^{17}$O$_{g.s.}$+$^{65}$Ni$_{0.31}$
\end{tabular} \right \}$ &  1.8  &  2.6    & 2.0  & $30\%$ & $-10\%$  \\
                  & $\left. \begin{tabular}{@{\ }l@{}} 
$^{17}$O$_{g.s.}$+$^{65}$Ni$_{0.69}$ \\ 
$^{17}$O$_{0.87}$+$^{65}$Ni$_{g.s}$  \\ 
$^{17}$O$_{g.s.}$+$^{65}$Ni$_{1.02}$ \\
$^{17}$O$_{g.s.}$+$^{65}$Ni$_{1.14}$ \\
$^{17}$O$_{g.s.}$+$^{65}$Ni$_{1.27}$
\end{tabular} \right \}$ & 2.8 & 3.1 & 2.5 & $28\%$ & $12\%$ \\
\hline
\end{tabular}

\raggedright

$^{a}$defined as $\delta_{\textnormal{DWBA}} = \frac{\sigma_{\textnormal{DWBA}}-\sigma_{\textnormal{CRC}}}{\sigma_{\textnormal{CRC}}}$.

$^{b}$defined as $\delta_{\textnormal{exp}} = \frac{\sigma_{\textnormal{exp}}-\sigma_{\textnormal{CRC}}}{\sigma_{\textnormal{CRC}}}$.
\label{resultsAngle}
\end{table*}

In Table~\ref{resultsAngle} we list the $\sigma_{\textnormal{exp}}$ along with theoretical ones for DWBA and CRC ($\sigma_{DWBA}$ and $\sigma_{CRC}$, respectively) for one-neutron transfer leading to low-lying states in the $^{17}$O, $^{29}$Si and $^{65}$Ni nuclei. Differential cross sections were integrated from $8\degree$ to $24\degree$ for the $^{17}$O, from $5\degree$ to $25\degree$ for the $^{29}$Si and from $11\degree$ to $38\degree$ for the $^{65}$Ni. In the DWBA calculations we have considered only the ground states of projectile and target in the entrance partition. We also define the deviation of $\sigma_{\textnormal{exp}}$ and $\sigma_{\textnormal{DWBA}}$ relative to the $\sigma_{\textnormal{CRC}}$ ($\delta_{\textnormal{exp}}$ and $\delta_{\textnormal{DWBA}}$, respectively). 

In the $^{18}$O$+^{16}$O system, experimental and CRC angle-integrated cross sections exhibit a very good agreement. For the $^{17}$O$_{g.s.} + ^{17}$O$_{g.s.}$ channel, for instance, $\sigma_{exp} = 4.2$ mb and $\sigma_{CRC} = 4.5$ mb that corresponds to a relative deviation of $-7\%$ using the \emph{psdmod} interaction. This is within the estimated experimental uncertainty of $10\%$. The angle-integrated cross section obtained within the DWBA approach is $6.5 \textnormal{mb}$ using the same interaction, which is $44\%$ higher than the CRC. Similar values is observed in the other channels and also when we considered the \emph{zbm} interaction. Considering that experimental values are close to the CRC ones and the large values obtained from DWBA calculation there is an indication that population of states in $^{17}$O through one-neutron transfer is affected by second-order processes. 

The comparison between angle-integrated cross sections for the $^{18}$O$+^{28}$Si system brings an alternative perspective about the effects of second-order processes in the one-neutron transfer. Although quality of the agreement between theory and experimental data is limited for the $^{17}$O$_{g.s.}+^{29}$Si$_{1.27}$ at large angles (see Fig.~\ref{results_si28}b) we note that relative deviation of $\sigma_{DWBA}$ is about $15\%$. The CRC calculation with SA from \emph{psdmwkpn} interaction reproduces the $\sigma_{exp}$ with deviation of $13\%$ and $16\%$ for the $^{17}$O$_{g.s.}+^{29}$Si$_{g.s.}$ and $^{17}$O$_{0.87}$+$29Si_{g.s.}$, respectively. The overall good agreement between $\sigma_{exp}$, $\sigma_{DWBA}$ and $\sigma_{CRC}$ obtained in these two reaction channels indicates that the one-neutron transfer to $^{28}$Si proceeds mainly through direct routes connecting the initial and final states. Same conclusion is obtained from the analysis of $^{28}$Si(d,p) as well [41]. However, it is not clear that the same holds in the population of $^{29}Si_{1.27}$. 
For the $^{18}$O$+^{64}$Ni system, the values of $\sigma_{exp}$, $\sigma_{DWBA}$ and $\sigma_{CRC}$ exhibit absolute deviations as high as $30\%$ and in this case is difficult to draw conclusions due to the need to sum different unresolved states in $^{65}$Ni nucleus. Nevertheless, the deviation between DWBA cross sections and the experimental data seems to be more acentauted than between CRC and data.

Summarizing, one-neutron transfer reactions to $^{16}$O and possibly to $^{64}$Ni are affected by competing processes like target and/or projectile excitation that takes flux from the reaction channel considered. On the other hand, in $^{28}$Si nucleus, the ground-to-ground transfer seems to be a single-step process like in (d,p) reaction. 

\section{\label{conc}Conclusions}
  
We present new experimental data for one-neutron transfer to $^{16}$O, $^{28}$Si and $^{64}$Ni induced by the ($^{18}$O,$^{17}$O) reaction at E$_{\textnormal{lab}} = 84\textnormal{ MeV}$. Coupled reaction channel calculations, using spectroscopic amplitudes derived from shell model calculations, give a good description of the cross sections to low-lying states in $^{17}$O, $^{29}$Si and $^{65}$Ni. We emphasize that no adjustable parameters are included in the optical potential for the direct reaction calculations and the present results give a good support to sequential two-neutron transfer calculations performed in previous works for the same systems.

Deviations between one-step DWBA and CRC angle-integrated cross sections, using same spectroscopic amplitudes, indicate that second-order processes are important in the one-neutron transfer to $^{16}$O and $^{64}$Ni while is not so relevant in the ground-to-ground transfer in $^{28}$Si. Determination of spectroscopic amplitudes by means of fittings of DWBA calculations to experimental data is a risky procedure when the transfer can be preceded by inelastic excitation of the projectile/target. Further systematic studies is needed to reveal the key ingredients that induce multi-step processes in the one-neutron transfer reactions induced by heavy-ions.

\section*{Acknowledgment}
This project has received funding from the European Research Council (ERC) under the European Union’s Horizon 2020 research and innovation programme (grant agreement No 714625).The Brazilian authors acknowledgment partial financial support from CNPq, FAPERJ and CAPES and from INCT-FNA (Instituto Nacional de Ci\^ {e}ncia e Tecnologia- F\' isica Nuclear e Aplica\c {c}\~ {o}es).  

\pagebreak

\end{document}